\documentclass[%
,aps%
,prl,showpacs,showkeys
,twocolumn
 ,secnumarabic%
,amssymb, amsmath,nobibnotes,floatfix]{revtex4-1}
\usepackage{docs}%
\usepackage{bm}%
\expandafter\ifx\csname package@font\endcsname\relax\else
 \expandafter\expandafter
 \expandafter\usepackage
 \expandafter\expandafter
 \expandafter{\csname package@font\endcsname}%
\fi

    \usepackage{latexsym}
    \usepackage[dvips]{color}
   \usepackage{slashed}
    \usepackage[dvips]{graphicx}
    \usepackage{lineno,hyperref}
\begin{document}

\def\p{\partial}
\def\f{\frac}
\def\A{\rm{A}}
\def\F{\rm{F}}
\def\cF{\cal{F}}
\def\rN{\rm{N}}
\def\P{\Pi}
\def\c{\phi}
%
\catcode`@=11
\catcode`@=11

\title{
\bf Lorentz symmetry violating low energy dispersion relations from a
dimension-five photon scalar mixing operator \footnote{Error!}}
\author{\bf{Avijit K. Ganguly}}
\email{avijitk@hotmail.com}
\affiliation{ Dept. Of Physics, MMV, Benaras Hindu University, Varanasi 221005, India.}
\author{ Manoj K. Jaiswal}
\email{ manojau@gmail.com}
\affiliation{Dept. Of Physics, MMV, Benaras Hindu University, Varanasi 221005, India.}

\date{April 2012}
\begin{abstract}
\noindent
Dimension-5 photon ($\gamma$) scalar ($\phi$) interaction terms usually appear 
in the bosonic sector of unified theories of electromagnetism 
and gravity. In these theories the three propagation eigenstates are different
from the three field eigenstates. Dispersion relation, in an external magnetic
field shows, that for a non-zero energy ($\omega$) out of the three
propagating eigenstate one has superluminal phase velocity $v_p$. 
During propagation, another eigenstate undergoes amplification or attenuation, 
showing signs of an unstable system. The remaining one maintains causality. 
In this work, using techniques from optics as well as gravity, we 
identify the energy ($\omega$) interval outside which $v_p \le c$ for the 
field eigenstates $~|\gamma_{\parallel}>$ and $ |\phi>$, 
and stability of the system is restored. The behavior of group velocity
$v_g$, is also explored in the same context. We conclude by pointing out its 
possible astrophysical implications.


\end{abstract}
\noindent

\pacs{11.25.Mj, 04.50.+h, 11.10Lm}
\keywords{Instability, Lorentz Violation}
\maketitle

\renewcommand{\thefootnote}{(\arabic{footnote})}



\noindent 
Scalar $\phi(x)$ photon $\gamma$ interaction through dimension five operator
originates in many theories beyond the standard model of particle physics, 
usually in the unified theories of electromagnetism and gravity \cite{duff}.
The scalars involved can be moduli fields of string theory, Kaluza-Klein 
(KK) particles from extra dimension, scalar component of the gravitational 
multiplet in extended supergravity models {\it etc.}, to name a few 
[\cite{Schmutzer}-\cite{Fischbach}].\\
\indent
Usually these models predict optical activity where the vacuum is turned
into a birefringent and  dichroic one\cite{raffelt}. 
As a polarized light beam passes through such a medium, its plane of 
polarization gets rotated. This particular aspect has been explored and exploited 
extensively in the literature to explain and predict many interesting physical 
phenomena 
\cite{Jain}.\\
\indent
In this work we point out other interesting aspects of such 
interactions encountered in the low energy sector of the theory, in a magnetized
background of field strength $\cal{B}$. 
The theory under consideration has a tree-level 
interaction term $g_{\phi\gamma\gamma}\phi F_{\mu\nu}F^{\mu\nu}$, where $g_{\phi\gamma\gamma}$ is a dimensionful coupling constant
between $\phi$ and Electro Magnetic (EM) field. This term is Lorentz Invariant (LI) 
and remains invariant under charge conjugation ({\bf C}), parity transformation 
({\bf P}) and time reversal ({\bf T}) symmetry transformations. 
Renormalizability of the theory is compromised because of the presence of 
dimensionful coupling constant $g_{\phi\gamma\gamma}$. However, in the presence of an 
external background magnetic field, all the good (both continuous and discrete
(i.e.~LI and CPT)) symmetries of the theory get compromised. Since 
theories violating CPT are known to violate Lorentz invariance \cite{greenberg}
hence causality; therefore, the explicit violation of both in a nontrivial 
background introduces modifications to the dispersion relations affecting 
phase velocity ($v_p$) and group velocity ($v_g$). The same also introduces
presence of unstable modes in a certain energy ($\omega$) domain. 
Some of these issues are explored below.\\
\indent
The theory under consideration has three {\it propagation 
eigenstates}, a scalar $|\phi\rangle$ and two transversely polarized photons $|\gamma_{||,\perp} \rangle $.
\!\! One of the eigenstate, $|\gamma_{||} \rangle $ has polarization vector parallel and the other one, $| \gamma_{\perp} \rangle $  has the same orthogonal 
to the background magnetic field $\cal{B}$.\\
\indent
 We point out in this note the two interesting possibilities
that may emerge 
from the solutions of the two eigenstates, $|\phi \rangle $ and  
$ |\gamma_{\parallel} \rangle$, (i) their phase velocity may become 
superluminal, (ii) their respective amplitudes may undergo attenuation or 
amplification, provided their energies lie in a certain interval. Within 
this energy interval the amplitudes of $|\gamma_{\parallel} \rangle$ 
and $|\phi \rangle$ may get amplified or  damped, thus they are 
non-propagating modes.\\
\indent
Naively, though this phenomena seem to get ameliorated, only 
at energy $\omega  =   \infty$, but through a careful analysis we show 
that there exists a finite energy interval outside which, the individual 
field states $|\gamma_{\parallel} \rangle$ and $|\phi \rangle$ are cured 
of this malady. In other words outside that interval, the solutions of 
the same are well behaved as far as their stability and the magnitudes of 
the phase velocities (i.e., $v_p$ for $|\phi \rangle$ and $|\gamma_{\parallel} 
\rangle$, both ) are concerned.\\
\indent
The same however can not be endorsed for the group velocity $v_g$ for 
these two states. The same (i.e., $v_g$) for $|\phi \rangle$ and 
$ |\gamma_{\parallel} \rangle $ reach luminal limit at $\omega= \infty$ only.\\
\indent
 We note in the passing that, 
the phase velocities of  $|\phi\rangle \pm |\gamma_{\parallel} \rangle $ do 
exhibit velocity selection rules as had been discussed in \cite{shore}.\\
%
%
\indent
 The other eigenstate 
$|\gamma_{\perp} \rangle$, is free from any pathological problems. It posses a 
stable solution as well as causal group and phase velocities, i.e., $v_g\!= \!v_p\!=\!c, ~\forall x,t \mbox{~and~} \omega $.\\
\indent
In this article we analyze the issues involved, from three different angles, (a)
using differential geometric arguments involving the properties of a metric
($g^{\mu\nu}_{(eff)} (\omega)$ in our context ) related to the stability of a 
manifold, as used in the context of relativity, (b) analyzing the Dispersion
Relations (DR), and (c) by explicit evaluation of 
the phase velocities ($v_p$) from the solutions of
the eigenstates, $|\gamma_{||,\perp}\rangle$  and $| \phi \rangle $, 
using principles of optics 
\cite{born-wolf}.\\
\indent
A critical analysis of the DR actually conforms with the findings, obtained
from the stability analysis of the effective metric, 
$g^{\mu\nu}_{(eff)}(\omega)$. 
The interesting part however is, $v_p$ turns out to be 
complex exactly in the same energy domain, as is predicted 
from the stability analysis of $g^{\mu\nu}_{(eff)}(\omega)$ as well 
as the dispersion relations. 
This indicates the system to be in an unstable state in the 
 relevant energy domain. A detailed further analysis  
reveals that, outside this energy range,
some of the Lorentz Invariance Violating 
(LIV) pieces in the expression of $v_p$, cancel out giving a LI and 
causal result.~We conclude by pointing out the possible implications of 
this result in astrophysical or cosmological contexts.\\  

\noindent
{\bf{Equations of motion:}}
%
The action for coupled scalar photon system, in 
four dimensional flat space, is given by:
\begin{eqnarray}
\!\!\!\!\!\!S \!=\!\! \!\int \!\!\!d^4x \! \!\left[\frac{1}{2}\partial{_\mu} \phi \partial^{\mu}\phi \! 
-\! \frac{1}{4}g_{\phi\gamma\gamma} \phi F_{\mu\nu}F^{\mu\nu}  
\! -\! \frac{1}{4}F_{\mu\nu}F^{\mu\nu} 
 \!\right].
\label{action1}
\end{eqnarray}
\indent
The equations of motion can be obtained from eqn.[\ref{action1}] by employing 
the usual variational principles. However, in what follows, we would rewrite 
eqn.[\ref{action1}] by decomposing the EM field tensor $F_{\mu\nu}$ 
into two parts, a slowly varying Background Mean Field 
${\bar{F}_{\mu\nu}}$, and an 
infinitesimal fluctuation $f_{\mu\nu}$ (i.e.,\!\!\! :\!\!\!
$F_{\mu\nu}={\bar{F}_{\mu\nu}}+f_{\mu\nu}$),
and then derive the equations of motion from the modified action. And without
loss of generality, we would consider a local inertial frame, where,
the only nonzero component of $\bar{F}^{\mu\nu}$, is $\bar{F}^{12}={\cal{B}} $. 
\!\!Assuming the magnitude of the scalar field to be of the order of the fluctuating EM field $f_{\mu\nu}$, one can linearize the resulting equations.  
The resulting equations of motion for the EM and the scalar fields 
turn out to be,
%
\vspace{-0.25em}
\begin{eqnarray}
 \partial_{\mu}f^{\mu\nu} 
=-g_{\phi\gamma\gamma} \partial_{\mu}\phi\bar{F}^{\mu\nu}, 
\label{emeq1}
 \\ 
\partial_{\mu}\partial^{\mu} \phi  =- \frac{1}{2} g_{\c\gamma\gamma} \bar{F}^{\mu\nu}f_{\mu\nu}.
\label{emeq2}
\end{eqnarray}
\noindent
Equation [\ref{emeq1}] describes the evolution of the two degrees of freedom
associated with the gauge fields and eqn. [\ref{emeq2}] 
describes the same for the scalar field.
Since eqn. [\ref{emeq1}] provides four equations for 
two degrees of freedom of the gauge fields, so one has 
to get rid of the extra relations by fixing a gauge
and using the constraint equation.\\
\indent
However, there is another 
way, i.e., by working in terms of the field strength tensors and making use of 
the Bianchi identity. In this work we will follow the second method. We will 
start with the Bianchi identity
$\partial_{\mu}f_{\nu\lambda}+\partial_{\nu}f_{\lambda\mu}+\partial_{\lambda}f_{\mu\nu}=0$ and 
multiply the same by $\bar{F}^{\nu\lambda}$; after this we operate 
$\partial^{\mu}$ 
on the resulting expression, to arrive at:
\vskip -0.3cm
%
\begin{eqnarray}
\partial_{\mu}\partial^{\mu} (f_{\lambda\rho} \bar{F}^{\lambda\rho})=-2\partial^{\lambda}\partial_{\mu}(f^{\mu\rho}
\bar{F}_{\rho\lambda}).
\label{bianchi1}
\end{eqnarray}
\noindent
Next we can multiply eqn.[\ref{emeq1}] by $ {\bar{F}}_{\nu\lambda}$, and subsequently operate ${\partial}^{\lambda}$ on the same to obtain, 
$$ \partial^{\lambda} \partial_{\mu} (f^{\mu\nu}{\bar{F}}_{\nu\lambda})=-g_{\phi\gamma\gamma}\partial^{\lambda} \partial_{\mu}\phi {\bar{F}}^{\mu\nu} 
{\bar{F}}_{\nu\lambda}.$$ 
Now using the relation given by eqn.[\ref{bianchi1}], on the last equation, 
we find the equation for the eigenstate $| \gamma_{||} \rangle$, given by:
\def\pk{k}
\begin{eqnarray}
\p_{\mu} \p^{\mu} \left(f_{\rho\sigma} \bar{F}^{\rho\sigma}/2\right) = g_{\c\gamma\gamma}
\p^{\lambda}\p_{\alpha}\phi (\bar{F}^{\alpha\nu}\bar{F}_{\nu\lambda}).
\label{eom1a}
\end{eqnarray} 
%
\noindent
The equation for the eigenstate $| \gamma_{\perp}\rangle$, can be obtained
by performing the same steps leading to eqn.[\ref{eom1a}], except
the multiplication of eqn.[\ref{emeq1}] by the factor $\bar{F}_{\nu\lambda}$. 
In this step, instead of $\bar{F}_{\nu\lambda}$ we have to use the 
multiplicative factor $\tilde{\bar{F}}_{\nu\lambda}$. This would lead us to:
\begin{eqnarray}
\p_{\mu}\p^{\mu}(f_{\nu\lambda}\tilde{\bar{F}}^{\nu\lambda}/2)=0,
\label{eom1b}
\end{eqnarray}
\indent 
It is easy to perform a {\it consistency check} on eqn.[\ref{eom1b}] 
using eqn.[\ref{eom1a}]. If we replace $\bar{F}_{\nu\lambda}$ by 
${\tilde{\bar{F}}}_{\nu\lambda}$ in eqn.[\ref{eom1a}] then we 
immediately recover eqn.[\ref{eom1b}], because the right hand side of 
eqn.[\ref{eom1a}] vanishes; since $\bar{F}^{\alpha\nu}\tilde{\bar{F}}_{\nu\lambda}=0$, because of our assumption that, for the background EM field, only 
$\bar{F}^{12} \ne 0$. Hence eqn.[\ref{eom1b}] is consistent.\\ 
\noindent
Now we introduce the new set of variables, 
$\psi \!=\!\frac{f_{\nu\lambda}\bar{F}^{\nu\lambda}}{2}$  and 
$\tilde{\psi}\!=\!\frac{f_{\nu\lambda}\tilde{\bar{F}}^{\nu\lambda}}{2}$, and use them in 
eqns.[\ref{eom1a},\ref{eom1b}], and subsequently go to momentum space, 
to obtain the dispersion relations. Those are given by:
%
\begin{widetext}
\begin{eqnarray}
 k^2\psi - g_{\c\gamma\gamma}
 \left(\pk_{\alpha}\bar{F}^{\alpha\nu}\bar{F}_{\nu\lambda}\pk^{\lambda}\right)\phi =0
,~~
 k^2 \tilde{\psi} =0~~\mbox{~and~}~~
k^2\phi - g_{\c\gamma\gamma}\psi=0.
\label{eomphi}
\end{eqnarray}
\end{widetext}
\noindent
Since we have assumed, that only $\bar{F}^{12}\!\!=\!\!{\cal{B}}\!\!=\!\!\tilde{\bar{F}}^{03}\!\!\ne\!\! 0$, then it follows from there, that:\!\!\!  
\def\cb{\cal{B}}
$\left(\bar{F}^{\mu\nu}\bar{F}_{\mu\nu} \right) \!\!= \!\!2{\cb}^2$
 and  
 $\left(\pk_{\alpha}\bar{F}^{\alpha\nu}\bar{F}_{\nu\lambda}\pk^{\lambda}
\!\right)\!\!=\!\!k^2_{\perp}\!{\cb}^2 $.\hspace{-0.1em} 
%
%
\noindent
Furthermore, if the angle between ${\cal{B}}$ and $\vec{K}$ is $\Theta$, then
the component of ${\vec K}$ normal to $\cal{B}$ is: $\vec{k}_{\perp}= \vec{K} 
\rm{sin}~\Theta$. Hence, using the same one can denote:
\vspace{-0.5em}
\begin{eqnarray}
 k^2_{\perp}{\cb}^2 = K^2 sin^2 \Theta {\cb}^2 \simeq \left(\omega {\cb} sin\Theta 
\right)^2 . 
\label{kpetc}
\end{eqnarray}
\vspace{-0.3em}
\noindent
While rewriting the eqn. [\ref{kpetc}], it was assumed that, $\omega \simeq K$ 
to zeroth order in the coupling constant $g_{\c\gamma\gamma}$. From now on, 
for the sake of brevity, we may denote ${\cb}sin\Theta={\cb}_T$,
at times.\\
\indent
In order to make the mass dimension of 
$\psi, {\tilde{\psi}}$ and $\phi$ same, we can multiply eqn.[\ref{eomphi}] by 
$\omega {\cal{B}} sin\Theta$ and redefine 
$\Phi =\omega {\cal{B}} sin\Theta \phi $. Upon doing the same, 
the coupled dispersion relations can 
be cast as a matrix equation:
%
\vskip -0.5cm
\begin{eqnarray} 
\left[ 
  \begin{matrix}
     & k^2          & 0                                                                    & 0                                                          \cr
     & 0          & k^2                                                                    &-g_{\c\gamma\gamma}\left(\omega {\cb}_T\right)     \cr
     & 0          &-g_{\c\gamma\gamma}\left(\omega{\cb}_T \right)                & k^2 
  \end{matrix}
\right] \!\!\!
\left[
\begin{matrix}
{\tilde{\psi}} \cr
\psi \cr  
\Phi
\end{matrix}
\right]= 0.
\label{m2}
\end{eqnarray}
\noindent
The real symmetric matrix, in eqn.[\ref{m2}], can be diagonalized 
by a orthogonal rotation through angle $\theta$, in the $\psi$-$\Phi$ plane.\\ 

\noindent
{\bf{Propagation Eigenstates: }} 
We already have explained, that, ${\tilde\psi}$ and $\psi$ have 
their respective polarization vectors $\perp$ and $\parallel$ to ${\cal{B}}$.
The off-diagonal elements in eqn. [\ref{m2}] make $\psi$ and 
$\Phi$ to mix during their space-time evolution; while 
$\tilde{\psi}$ remains unaffected. Next we  diagonalize 
eqn.[\ref{m2}], by the orthogonal transformation discussed before, 
and express the same as:
%
\begin{eqnarray}
\!\!\!\!\left(\begin{matrix}
     &\!\!\! k^2     &\!\!\!0   &\!\!\! 0  \cr
&\!\!\!\!0     &\!\! k^2-g_{\c\gamma\gamma}{\cb}_T \omega   &\!\!\!  0 \cr
&\!\!\!\!0   & \!\!\! 0  &\!\!\!\!\!\!\!\!\!\! k^2+g_{\c\gamma\gamma }{\cb}_T\omega
\end{matrix}
\right)\!\!\!
\left(
\begin{matrix}
\tilde{\psi} \cr
\frac{ \Phi+\psi }{\sqrt{2}} \cr
\frac{\Phi -\psi}{\sqrt{2}}
 \end{matrix}
\right)\!\!\!=\!0.
\label{diagbasis1}
\end{eqnarray}
%
\noindent
It is easy to see from eqn. [\ref{diagbasis1}] that, the {\it
propagating eigenstates}
$\tilde{\psi}$, $\frac{\Phi+\psi}{\sqrt{2}}$ and
$\frac{\Phi-\psi}{\sqrt{2}} $ satisfy the following dispersion 
relations,
\begin{eqnarray} 
\omega &=& K  
\label{livs}  \\ 
\omega_{+} &=& \pm \sqrt{K^2+g_{\c\gamma\gamma }{\cb}_T\omega }  
\label{livas} \\
\omega_{-} &=& \pm \sqrt{K^2-g_{\c\gamma\gamma }{\cb}_T\omega}.
\label{livsn}
\end{eqnarray}
%
\noindent
We point out that the dispersion relations obtained from eqns.[\ref{m2},\ref{diagbasis1}] are identical to those obtained in [\cite{Iacopini,Miani,Ganguly3}], provided appropriate limits are taken.\\
\indent
Upon dividing eqns.[\ref{livas},\ref{livsn}] by $K$, we arrive at the 
expressions for the phase velocities, $v^{\pm}_{p}={\sqrt{1\pm g_{\phi\gamma\gamma}\cal{B}_{T}/\omega }}$, corresponding to the propagation eigenstates, $[\frac{\Phi\pm\Psi}{\sqrt{2}}]$.  It is easy to verify that, for 
$g_{\phi\gamma\gamma}\cal{B}_{T} > \omega$, the magnitude of $v^{+}_{p}>1$ 
that is, phase velocity of the eigenstate $[\frac{\Phi+\Psi}{\sqrt{2}}]$ 
propagates with superluminal speed, and $v^{-}_{p}$ is complex so the amplitude
of the corresponding eigenstate $[\frac{\Phi+\Psi}{\sqrt{2}}]$ would be 
attenuated or damped, {\it as was mentioned in the beginning.}\\
%

\noindent
{\bf{Effective metric:}}
To understand more about the Lorentz Invariance Violating (LIV) dispersion
relation in the magnetized vacuum for the mixed propagation eigenstate $\left[\frac{\Psi+\Phi}{\sqrt{2}}\right]$, we note that the dispersion relation 
for the same can be written as $g^{\mu\nu}_{(eff)}(\omega)k_{\mu}k_{\nu}=0$, where,
$ g^{\mu\nu}_{(eff)}(\omega)= diag \left( \left[ 1-\frac{g_{\c\gamma\gamma }{\cal{B}} sin\Theta}{\omega} \right], -1,-1,-1 \right)$ and $k_{\mu}$ is the usual  wave 4-vector. The form of the effective metric 
given above is similar to the ones discussed in the context 
of Doubly Special Relativity (DSR) \cite{kimberly}. We clarify at the outset
that the same has been obtained, here, by demanding that the dispersion 
relation can be written as a quadratic of $k_{\mu}$'s, like the same for 
mass-less particles.
\footnotetext{The appearance of energy $\omega$, is the reflection of the fact 
that the dispersion relation itself does not respect Lorentz symmetry, because 
the external magnetic field in the system. However as would be shown that 
in-spite of breaking of Lorentz violation by Magnetic field, the same is 
restored for modes with $\omega < \omega_c $.}
One may interpret this effective metric as, the metric of the 
underlying spacetime over which $\left[\frac{\Psi+\Phi}{\sqrt{2}}\right]$ 
is propagating. The inverse of the same is $g_{\mu\nu (eff)}(\omega)$ is given by,
$ g_{\mu\nu (eff)}(\omega) = diag \left(
     \frac{1}{\left[1-\frac{g_{\c\gamma\gamma } {\cal{B}} sin\Theta}{\omega}\right]}, -1, -1,-1 \right)$. 
Next we would perform stability analysis of the system using this metric.\\

\noindent
{\bf{Stability Analysis Using $\mathbf{g_{\mu\nu (eff)}(\omega)}$:}}
It has been pointed out in \cite{LL}, that, for a space-time to be  
stable, the determinant of it's metric must be negative, else the system
is unstable and would decay to a stable ground state. The purpose of 
writing the effective metric was to find out if there exists a bound or 
interval over which determinant of the same is negative indicating 
possibility of attenuation or growth of the amplitudes of the 
eigenmodes.\\
\indent
If we take a critical look at $g_{\mu\nu (eff)} (\omega)$, it is clearly seen that unless
$ \omega > g_{\c\gamma\gamma } {\cal{B}} sin\Theta = \omega_c $  the value of 
$Det |g_{\mu\nu (eff)} (\omega)|> 0$, hence there would be growth (instability)  
or damping (attenuation) in the system. Now if we  go back to eqn. 
[\ref{livas}] one can verify that, the same can be  recast in the 
following form,
$\left( \omega - \frac{ g_{\c\gamma\gamma } {\cal{B}} sin\Theta}{2}\right)^2-
 \left( \frac{ g_{\c\gamma\gamma } {\cal{B}} sin\Theta}{2} \right)^2 = K^2 $. 
Accordingly, for $\omega < \omega_c $, wave vector $K$ becomes imaginary, 
signaling attenuation or growth of amplitude. Therefore, we are tempted to conclude that the deductions of \cite{LL}, holds even for the effective 
metric $g_{\mu\nu(eff)}(\omega)$.\\

\indent
{\bf{Causal Stability:}}
It has been pointed out in \cite{hawking,astro-2,novello} that the
stability of causal manifolds are governed by two conditions, (a) the 
underlying metric has
to be Lorenzian and (b) there should  exist a scalar time-like function 
$ T(x) $, i.e. continuous and infinitely differentiable every where 
on the manifold; and covariant derivative of $T(x)$ i.e., $D_{\mu}T(x)\ne 0$, 
and $g^{\mu\nu}_{(eff)}(\omega)D_{\mu}T(x)D_{\nu}T(x) >0$ \cite{Klinkhammer, klinkhammer2}.  In our case  both the conditions are satisfied, provided, we 
take $T(x)=t$ as the time coordinate (i.e., illustrating the absence of closed 
time like or space like curves).\\
%
\vskip .5cm
\noindent
{\bf{Inhomogeneous Wave Equations:}}
\noindent
It is possible to get the solutions for the {\it propagating eigenstates} 
$\tilde{\psi}$, $\frac{\Phi+\psi}{\sqrt{2}}$ and $\frac{\Phi-\psi}{\sqrt{2}}$  
from the dispersion relations given by 
eqns.[\ref{livs},\ref{livas},\ref{livsn}], that follows 
from eqn.[\ref{diagbasis1}].\\  
\indent
Sometimes, presenting results in its full generality, becomes a 
fruitful and instructive exercise in many areas of exact science. 
It helps in pointing out potential sources of new scientific features.
 Keeping this philosophy in mind we express the solutions of the 
coupled set of equations, as an explicit  function of the 
rotation angle $\theta$, in the $\psi -\Phi$ plane. They have the 
following form, 
\vskip -0.13cm
\begin{eqnarray}
\left(
\begin{matrix}
\tilde{\psi} \cr
cos \theta ~\psi + sin \theta ~\Phi \cr
-sin \theta ~\psi + cos\theta ~\Phi
 \end{matrix}
\right)
=
\left(
\begin{matrix}
A_{0} e^{i\left( \omega t - k.x \right)} \cr
A_{1} e^{i\left( \omega_{+} t - k.x \right)} \cr
A_{2} e^{i\left( \omega_{-} t - k.x \right)} 
\end{matrix}
\right)
\label{solns}
\end{eqnarray}
\vskip -1.25em
\noindent
It is not difficult to see, that for $\theta=\pi/4$, one recovers,
back the expressions for propagating eigen states,  $\frac{\Phi \pm \psi}{\sqrt{2}}$.    
We would like to mention here that, we would not consider, 
$\theta=\frac{\pi}{4}$, till we reach an appropriate point.\\
\indent
The constants, $A_{0},~A_{1}$~ and $A_{2}$ appearing in eqn.[\ref{solns}],
are to be derived from the boundary conditions one imposes on the 
dynamical degrees of freedom. The solutions for the dynamical variables, 
from eqn.[\ref{solns}], turn out to be,
\vspace{-1.21em}
\begin{widetext}
\begin{eqnarray}
\!\!\! \tilde{\psi} (t, x)\! = \!\!A_{0} e^{i\left( \omega t - k.x \right)}, 
\psi (t,x)\!\! =\! \!\left[A _{1}cos \theta  e^{i\omega_{+} t}\! 
-\!\! 
 A _{2} sin \theta  e^{i\omega_{-} t}\right]\!e^{-ikx} \mbox{~and~}
\Phi (t,x) \!=\!\!\left[ A _{1} sin \theta  e^{i\omega_{+} t}\! + \!
 A _{2}  cos \theta  e^{i\omega_{-} t}\! \right]\!e^{- ik.x}\!.
\label{solnforfields}
\end{eqnarray}
\end{widetext}
\vspace{-1.2em}
\noindent
In the following we consider the boundary conditions, $\Phi (0,0) = 0$
and $ \psi(0,0) =1 $. With these boundary conditions,  we have, 
$ \frac{A_{2}}{sin \theta}= -1 $ and  the solution for $\psi$ turns out to be,
\begin{eqnarray}
\!\!\!\! \psi(t, x)\!\!=\! 
\left[ cos^2 \theta \,\, e^{i\left( \omega_{+} t - k.x \right)}\!\! +\!\! 
  sin^2 \theta \,\,  e^{i\left( \omega_{-} t - k.x \right)} \right].
\label{solnpsi}
\end{eqnarray}
\noindent
Defining, $ a^2_{x} (t)=\left({\rm{\cal{R}}}e \left[\psi (t,0)\right]\right)^2 
+ \left(\rm{\cal{I}}m \left[\psi (t,0) \right] \right)^2$, we get the 
following form for $\psi \left(t,x \right)$,
%
\begin{eqnarray}
\!\!\!\psi(t,x)\!\! =\!\! a_{x} (t) e^{\! i \!\left( tan^{-1} \! \left[\frac{ cos^2 \theta  sin \omega_{+} t+ sin^2 \theta \, sin \omega_{-} t }{  cos^2 \theta \, cos \omega_{+} t+ sin^2 \theta \, 
cos \omega_{-} t} \right]- k x\!\! \right)}
\label{inhomosoln}
\end{eqnarray}
%
\noindent
A wave equation of this type is usually called inhomogeneous wave equation 
\cite{born-wolf}. The phase velocity for such a system, where the solution 
is represented by, $a(t)e^{i(\varphi(t) - k.x) }$ is defined by,
$v_p = \frac{1}{K}\frac{\partial \varphi(t)}{\partial t}$.\\
%
\indent
In more complicated physical situations, when medium effects, 
polarization effects due to strong external fields etc., 
are taken into account, 
the angle $\theta$ would depend on those parameters. Hence $\varphi(t)$ may become a 
complicated function of time.
As a result, the phase velocity, $v_p$ may become a function of time 
with varied physical implications.\\  
\indent
However, for the simple case in hand, substituting $\theta=\frac{\pi}{4}$ 
in eqn. [\ref{inhomosoln}], followed by some algebra, it is easy
to demonstrate that, $\varphi(t)=\frac{\left(\omega_++\omega_-\right)t}{2}$. 
Now using the same in the expression for phase velocity $v_p$  yields,
\begin{eqnarray}
v_p = \left(\frac{ \omega_{+} + \omega_{-}}{2K}\right)
\label{vfot}
\end{eqnarray} 
%
\noindent
Using eqns. [\ref{livas}] and [\ref{livsn}] in eqn.[\ref{vfot}] and 
considering the dispersion relation to zeroth order in 
$g_{\phi\gamma\gamma}$, i.e., $\omega \simeq K$, we obtain, 
%
%
\begin{eqnarray}
v_p\! = \!\frac{1}{2}\!\! \left[\!  
\sqrt{\!\!\left(\!\!1\!\! - \frac{g_{\c\gamma\gamma} \cal{B}_{T}}{\omega}\!\! \right)}
\!+\!\sqrt{\!\! \left(\!\! 1\!\! + \! \frac{g_{\c\gamma\gamma}\cal{B}_{T}}{\omega}
 \right) }
                                             \right]
\label{phvel}
\end{eqnarray}
%
\noindent
The expression for phase velocity, as given by eqn. [\ref{phvel}], provides an 
interesting limit for $\omega$; in order to have a  real phase velocity, one 
must have $\omega \ge g_{\c\gamma\gamma} \cal{B}_{T} $. So in principle 
one can define an expansion parameter $\delta=\frac{g_{\c\gamma\gamma}\cal{B}_{T} }{\omega}$, and  perform an all order expansion of 
$\left(1 \pm \frac{g_{\c\gamma\gamma} \cal{B}_{T}}{\omega}\right)^{1/2}$, 
in powers of $\delta$, for $\delta <1$, and be 
convinced that the magnitude of $v_p$ stays less than $c$,
i.e. phase velocity is causal.\\ 

\noindent{\bf{Group velocity:}}
Group velocity for the situation under consideration is given by, $v_{g}=|\frac{\partial\dot{\varphi}}{\partial K}|$. Using the expression for $\dot{\varphi}$, in the last relation, we obtain the expression for group velocity in terms 
of $\delta$,
\begin{eqnarray}
\hskip -1cm v_g\!=\!\frac{1}{2}\!\left[\!\frac{1-\frac{\delta}{2}}{\sqrt{1-\delta}}+
\frac{1+\frac{\delta}{2}}{\sqrt{1+\delta}}\!\right]\!,  
\label{grvel}
\end{eqnarray} 
\indent
Expanding the right-hand side of eqn.(\ref{grvel}) in powers of $\delta$ 
(assuming $\delta < 1$), 
one finds that $v_g >1$, even when, $ 0<\delta<1$. Of~course the problem  
of having complex $v_g$ is avoided by considering $\delta <1$, however the issue 
of superluminality remains. We believe, that this is an artifact of the special background that violates 
Lorentz and CPT invariance. The presence of this special background may
be responsible for making $v_g$ of the $|\gamma_{\parallel} \rangle$ state 
superluminal.
\vspace{-0.22em}
\begin{figure}[ht!]
\hskip .15cm
\includegraphics[scale=0.9]{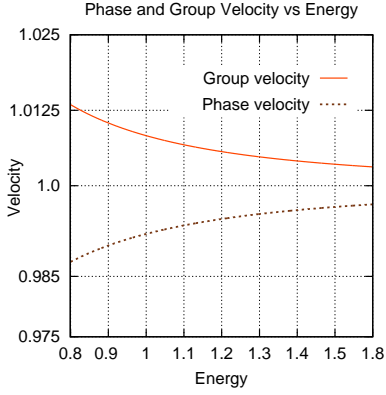}
\vskip -.40cm
\caption{\label{fig1}\small{Plot of Velocity vs Energy. Energy is plotted in units of $10^{-15}$ GeV. The other parameters being $g_{\c\gamma\gamma}=(10^{-11} \rm{GeV})^{-1}$ and 
${\cal B}$=$10^{9}$ Gauss.}}
\label{fig:1}
\end{figure}
\noindent
In Fig.[\ref{fig:1}], we have plotted $v_g$ and $v_p$, for various 
values of $\omega$. As can be seen from the plots, that as energy, $\omega \to \infty$  the group (phase) velocity, $v_{g(p)} \to 1$.\\
\indent
 The solution for $|\phi \rangle $, is similar to $|\gamma_{\parallel} \rangle$ 
modulo a constant phase factor. It seems that, for $\omega  < \omega_c$, there 
is energy exchange between these two modes. A detailed understanding of the physics of energy transfer as well as how the system behaves once the 
back-reaction of 
the propagating modes on the background is taken into account,
following [\cite{cooper} and \cite{boyanovsky}], seems to be an important issue. However addressing the same are beyond the aim and the scope of the current article and would not be dealt with any further here.\\  

\noindent
{\bf{Signature:}}
 In astrophysical situations synchrotron or curvature radiation
is the most common process of non-thermal emission. As is well known, from \cite{Ginzburg},  that such 
radiations are always {\it polarized} along and orthogonal to 
the ${\mathcal{B}}-V$ plane. Where $V$ is the instantaneous velocity vector  of the
radiating charged particle. The synchrotron amplitudes of the electromagnetic radiation 
for these two polarized states are given by,
%
\begin{eqnarray}
A_{\perp}= \frac{\sqrt{3}\gamma^2_{f}\theta_e}{\omega_r}\sqrt{1+
 \gamma^2_{f}\theta^2_e } K_{1/3}\left(\frac{\omega}{2\omega_r}\right), 
\nonumber \\
\vspace{-2em}
A_{\parallel}= i\frac{\sqrt{3}\gamma_{f}}{\omega_r}\left( 1+
 \gamma^2_{f}\theta^2_e \right) K_{2/3}\left(\frac{\omega}{2\omega_r}\right). 
\label{power}
\end{eqnarray}
 In eqns. [\ref{power}], $\gamma_f$ is the Lorentz factor, $\omega_r=\frac{3 \gamma^3_f}{\rho}$ is the cutoff frequency and $\rho$ is the radius of curvature of the 
trajectory of the radiating particle. Lastly  $\theta_c=\frac{1}{\gamma_f}$ is the opening angle of the radiating cone.\\ 
\indent
Since for $\omega < \omega_c$, the only evolving polarized state is 
$|\gamma_{\perp} \rangle $ when dimension-5, $\phi F_{\mu\nu}F^{\mu\nu}$ 
interaction is present, therefore, to a far away observer, the synchrotron 
radiation would appear to be {\it linearly} polarized.\\
\indent 
So, the differential intensity spectrum per unit energy, per unit 
solid angle at the source for the $| \gamma_{\perp}\rangle$ state, following 
eqn.[\ref{power}], is given by:
%
$\frac{d^2 I}{d \omega d\Omega}= \frac{(e\omega)^2}{4\pi^2}
\left(|A_{\perp}|^2 \right)$. Further more, if all the 
astrophysical absorption mechanisms are negligible, then the 
magnitude of $\frac{d^2 I}{d \omega d\Omega}$, at the source,
as well as, at the observation point would remain the same.
Therefore, the differential intensity spectrum for two different energies 
$(\omega_{1}, \omega_{2} < \omega_c)$, would be related to the 
respective energies, $\omega_{1}$ and $\omega_{2}$ by,
\begin{eqnarray}
\!\!\!\!\!\!\!\!\!\frac{\frac{d^2I(\omega_1)}{d\omega_1 d\Omega}}
{\frac{d^2I(\omega_2)}{d\omega_2 d\Omega}}
\!\!=\!\!\left[\frac{\omega_1 K_{\frac{1}{3}}\!\! \left(\frac{\omega_1}{2\omega_c}\right)}
{\omega_2 K_{\frac{1}{3}}\!\!\left(\frac{\omega_2}{2\omega_c}\right) } 
\right]^2\!\!\!
\!\!\rm{~,implying,~~} \!\!\frac{\omega_2}{\omega_1}\!\!=\!\!
\left[\frac{\frac{d^2I(\omega_1)}{d\omega_1 d\Omega}}
{\frac{d^2I(\omega_2)}{d\omega_2 d\Omega}}\right]^{\frac{3}{4}}\!\!\! .
\label{freqint}
\end{eqnarray}
This is the {\it intensity -- energy} relation. While deriving the same
(i.e., eqn. [\ref{freqint}]), we have used eqn.[\ref{power}] and 
expanded $K_{1/3}(x)$ in decending powers of $x$.\\ 
\indent
Next we would like to relate this {\it intensity -- energy} relation (eqn. [\ref{freqint}]) with {\it the rotation measure}.\\
\indent
Since the intervening media between the source and the far-away  observer is 
magnetized and composed of nonrelativistic, degenerate electrons; the Plane Of 
Polarization (POP) of a polarized light (of energy $\omega$) passing through 
the same would undergo Faraday Rotation (FR), given by 
\cite{GKP}, 
\begin{eqnarray}
\varphi = \frac{\alpha \pi \left({\cal{B}} \cos{\Theta} \right)}{\omega^2 m_e}
nl  + \zeta.
\label{roa}
\end{eqnarray}
\noindent 
\noindent 
Here, $\zeta$ is the angle between POP and ${\hat{\cal B}}$ at source and $l$
is the length of the path travelled. Rest of the symbols in eqn. [\ref{roa}] have their 
usual meaning.\\ 
\indent 
Since the net rotation measure ($\varphi- \zeta$) due to FR goes as 
$\frac{1}{\omega^2}$; therefore, for a multi-frequency plane polarized 
light beam, the ratio of the two {\it rotation  measures} at two distinct
energies  ($\omega_1$ and $\omega_2$), will be given by the 
following relation,\\ 
%
\indent
\begin{eqnarray}
\omega_2/\omega_1=\sqrt{\left(\varphi_1-\zeta \right)/\left(\varphi_2-\zeta 
\right)}~.
\label{freqpol}
\end{eqnarray}
\noindent
Eqn.[\ref{freqpol}] may henceforth be termed as Energy-Dependent-Rotation 
Measure (EDRM).\\
\indent
 Now we can use eqns. [\ref{freqint}] and [\ref{freqpol}], to arrive at a 
relation between the rotation measure and the differential intensity spectrum,
for $\tilde{\psi}$ (~i.e.,
the solution for  $|\gamma_{\perp} \rangle$ state), and the same is: 
\begin{eqnarray}
\frac{\left(\varphi_1-\zeta \right)}{\left(\varphi_2-\zeta 
\right)}=\left[\frac{d^2I(\omega_1)}{d\omega_1 d\Omega} \div
\frac{d^2I(\omega_2)}{d\omega_2 d\Omega}\right]^{\frac{3}{2}}.
\label{roai}
\end{eqnarray}
\noindent
\noindent
For magnetic field strength at source, ${\cal{B}} \sim 10^{9}$ Gauss, 
and $g_{\phi\gamma \gamma} \sim \left( 10^{11} GeV\right)^{-1}$, we have 
$\omega_c \sim 10^{-5}eV$ which lies in the radio range.\\ 
\indent
So, the polarization versus (differential) intensity 
distribution pattern, for plane polarized light, 
in the energy range, $0 <\omega < 10^{-5} eV $, 
from distant astrophysical objects (with dominant synchrotron source), 
should behave according to eqns. [\ref{freqint}] and [\ref{roai}].\\

\indent 
Conversely, for $\omega$ above $\omega_c$, both, $\tilde\psi$ and 
$\psi$ would propagate in space-time. And $\psi$ would undergo 
amplitude modulation because of mixing with $\Phi$. Hence, the 
emerging light beam may bear some appropriate polarimetric
\cite{mybook} 
and dispersive signatures of $g_{\phi\gamma\gamma}\phi \bar{F}_{\mu\nu}F^{\mu\nu}$ interaction,  when the Faraday and the mixing effects are considered together,
 provided, the same is realized in nature.\\
\indent 
Similar signatures from different astrophysical radio sources were reported
in \cite{cfj} and \cite{TJ} sometimes back. They may have some 
implications for the situation we have discussed in this note. However one 
should work with the new data sets before coming 
to a definite conclusion.\\

\noindent
{\bf Acknowledgment:} We would like to thank the referee for his constructive criticism
and suggestions.   
\vspace{-2em}

\end{document}